\documentclass[prl,twocolumn,showpacs]{revtex4}
\usepackage{amsfonts,amsmath}

\begin{document}

\title  {Eigenvalue statistics of the real Ginibre ensemble}

\author{Peter J. Forrester$^{\dagger}$ and Taro Nagao${}^*$}

\affiliation
 {$^{\dagger}$Department of Mathematics and Statistics, University of Melbourne, 
Victoria 3010, Australia \\
${}^*$
Graduate School of Mathematics, Nagoya University, 
Chikusa-ku, Nagoya 464-8602, Japan}

\begin {abstract}
The real Ginibre ensemble consists of random $N \times N$ matrices formed from i.i.d.~standard
Gaussian entries. By using the method of skew orthogonal polynomials, the general $n$-point
correlations for the real eigenvalues, and for the complex eigenvalues, are given as
$n \times n$ Pfaffians with explicit entries. 
A computationally tractable formula for
the cumulative probability density of the largest
real eigenvalue is presented. This is relevant to May's stability analysis of biological webs.
\end{abstract}

\pacs{02.50.-r, 05.40.-a,
75.10.Nr}
 \maketitle

Dyson's three fold way \cite{Dy62c} is a viewpoint on the foundations of random matrix theory,
showing how consideration of time reversal symmetry leads to three classes of ensembles of relevance to
quantum mechanics. The three ensembles are catalogued by the classes  of unitary matrices which
leave the ensemble invariant --- orthogonal (time reversal symmetry is an involution),
unitary (no time reversal symmetry), and symplectic (time reversal symmetry is an anti-involution).
For an ensemble theory of Hermitian matrices, an equivalent characterization is that the matrix
elements be real, complex and real quaternion respectively. 

Both as a concept, and as a calculational tool, the three fold way has been highly successful.
As a concept, allowing for global symmetries in addition to that of time reversal gives a
classification of the former in terms of the ten infinite families of matrix Lie algebras
\cite{Zi97}. This classification 
now provides theoretical underpinning to fundamental phenomena in mesoscopic physics \cite{Be97},
disordered systems \cite{Ef97a}, and low energy QCD \cite{VW00}, in additional to the study of
the statistical properties of 
quantum spectra for which it was originally intended. A good deal of the success relates to the
matrix ensembles of the three fold way and its generalization
being exactly solvable --- analytic forms are available
for all key statistical quantities, allowing for quantitative theoretical predictions.

Soon after the formulation of the three fold way, Ginibre \cite{Gi65} presented as a mathematical
extension an analogous theory of non-Hermitian random matrices. The entries are taken to be 
either real, complex or real quaternion. Like their Hermitian counterparts,
it transpires that such random matrices have physical relevance.

Consider the complex case first. Then the joint eigenvalue probability density function (PDF) 
is proportional to
\begin{equation}\label{2.1}
\prod_{l=1}^N e^{-|z_l|^2} \prod_{j<k}^N |z_k - z_j|^2, \quad
z_j := r_j e^{i \theta_j}.
\end{equation}
This
can be recognised as the Boltzmann factor for the two-dimensional one-component plasma in
a disk, or the absolute value squared of the wave function for free fermions in a plane,
subject to a perpendicular magnetic field and confined to the lowest Landau level
\cite{Fo98a}. In the study of chaotic dissipative quantum systems, the statistical properties of
eigenvalues for certain model maps are well described by the corresponding statistical properties
implied by this PDF \cite{Ha90}.

In the case of real quaternion elements, the eigenvalues come in complex conjugate pairs.
The eigenvalue PDF of the eigenvalues in the upper half plane is
proportional to
\begin{equation}\label{2.2}
\prod_{l=1}^N e^{-2|z_l|^2} |z_l - \bar{z}_l|^2 
\prod_{j<k}^N |z_k - z_j|^2|z_k - \bar{z}_j|^2
\end{equation}
Up to an extra one body factor $\prod_{l=1}^N |z_j - \bar{z}_j|$,
the eigenvalue PDF of the eigenvalues in the upper half plane is
proportional to the
Boltzmann factor for the two-dimensional one-component plasma confined to a semi-disk,
bounded by a dielectric material of dielectric constant $\epsilon = 0$ along the straight
edge \cite{Sm82,Fo02}.

Both joint PDFs for the complex and real quaternion cases are contained in Ginibre's
paper \cite{Gi65}.
However, in the case of real elements, it wasn't until a further twenty-five or so years
later that the joint distribution was computed, first by Lehmann and Sommers \cite{LS91}, then
by Edelman \cite{Ed95}. Part of the difficulty is that the joint PDF is not absolutely continuous.
Rather, there is a non-zero probability that for $N$ even (odd) there will be an even (odd)
number of real eigenvalues for all even (odd) positive integers up to $N$. 
The final result is that for $k$ real eigenvalues ($k$ of the same parity as $N$), the joint PDF is
\begin{eqnarray}\label{3.1}
&&\! \! \! \!{1 \over 2^{N(N+1)/4} \prod_{l=1}^N \Gamma(l/2) }
{2^{(N-k)/2} \over k! ((N-k)/2)! } \nonumber \\
&& \!\!\!\! \times \Big | \Delta(\{\lambda_l\}_{l=1,\dots,k} \cup
\{ x_j \pm i y_j \}_{j=1,\dots,(N-k)/2}) \Big | \nonumber \\
&&  \!\!\!\! \times
e^{- \sum_{j=1}^k \lambda_j^2/2} e^{\sum_{j=1}^{(N-k)/2}(y_j^2 - x_j^2)}
\prod_{j=1}^{(N-k)/2} {\rm erfc}(\sqrt{2} y_j)
\end{eqnarray}
where $\Delta(\{z_p\}_{p=1,\dots,m}) := \prod_{j < l}^m (z_l - z_j)$. Here 
$\lambda_l \in (-\infty, \infty)$ while $(x_j,y_j) \in {\mathbb R}^2_+$,
${\mathbb R}^2_+ := \{ (x,y) \in {\mathbb R}^2 : \, y>0 \}$. Integrating (\ref{3.1}) over
$\{\lambda_l\} \cup \{x_j + i y_j \}$ gives the probability that there are precisely $k$
eigenvalues. The simplest case in this regard is when $k=N$ (i.e.~all eigenvalues real),
and it is found that the sought probability is equal to $2^{-N(N-1)/4}$ \cite{Ed95}. 
For $k=2$ an
evaluation in terms of a single definite integral has been given recently in
\cite{AK07}, while \cite{KA05} reduces the calculation for general $k$ down to
an expression of the same computational complexity as our Eq.~(\ref{10}) below.

Perhaps the first applied study to draw attention to the eigenvalues of random real matrices
was that of May \cite{Ma72a}, in the context of the stability of large biological webs.
The very general setting \cite{Mc75} is to consider an $N$-dimensional vector $\vec{x}(t)$
with components specified as the solution of the coupled first order system
${d x_i(t) / dt} = F_i(\vec{x}(t))$, $(i=1,\dots,N)$ for some nonlinear functions $F_i$.
Assuming an isolated equilibrium solution $\vec{x}^0$, linearization about this point leads to
the linear matrix differential equation
\begin{equation}\label{0.2}
{d \vec{y}(t) \over dt} = A \vec{y}(t)
\end{equation}
where $A$ is an $N \times N$ matrix.
The system is stable if all eigenvalues of $A$ have a negative real part. To model the effect of
random coupling between components on a stable equilibrium, the matrix $A$ is written
$A = - 1_N + B$ where $B$ is a dilute matrix (fraction $1-c$ of its elements zero) with mean zero and
variance $s^2$. The May criterion asserts that stability requires $s \sqrt{Nc} < 1$. Indeed in
the case $c=1$ this is consistent with limit theorems for the spectral radius of random real matrices
proved subsequently \cite{Ge86,Ba97}. Neural networks are further examples of complex
webs to which such a random matrix based stability analysis is relevant 
\cite{FJD06,AS06,RA06}. The results obtained below allow the evaluation of the
probability of stability in the borderline case of the May stability criterion,
\begin{equation}\label{0.5}
s \sqrt{Nc} = 1.
\end{equation}

As with the matrix ensembles of Dyson's three fold way, all correlations and a number
of key distributions for the complex  and real quaternion Ginibre ensembles
have been calculated exactly \cite{Me91,Fo02}. The Fourier transform of the two-point
correlation (structure function) is a quantity of key importance to the plasma and fermion
interpretation of (\ref{2.1}), while the decay of the two-point function along the boundaries
indicates general physical principles (non-zero dipole moment of the screening cloud in the
case of (\ref{2.1}); vanishing dipole moment for (\ref{2.2})). Further, the distribution function for the
spacing between eigenvalues in the bulk can be compared against data obtained from
dissipative maps \cite{Ha90}, while the density fluctuations in a large disk within the bulk
indicate further general physical principles \cite{Ma88}.

In contrast to the situation for (\ref{2.1}) and (\ref{2.2}), the correlations and distributions have
not in general been computed for
the real Ginibre ensemble. Exceptions are the density of real eigenvalues \cite{EKS94}
\begin{eqnarray}\label{6.1}
&&\rho_{(1)}^{\rm r}(x) =  {1 \over \sqrt{2\pi} } \Big (
{\Gamma(N-1,x^2) \over \Gamma(N-1)} 
\nonumber \\ && \qquad + {2^{N/2-3/2} \over \Gamma(N-1)} |x|^{N-1} e^{-x^2/2}
\gamma( {N-1 \over 2}, {x^2 \over 2} ) \Big )
\end{eqnarray}
with $\Gamma(p,x) := \int_x^\infty t^{p-1} e^{-t} \, dt$,
$\gamma(p,x) := \int_0^x t^{p-1} e^{-t} \, dt$, and the density of complex eigenvalues \cite{Ed95}
\begin{equation}\label{6.2}
\rho_{(1)}^{\rm c}((x,y)) = \sqrt{2 \over \pi} {\Gamma(N-1,x^2 + y^2) \over \Gamma(N-1)} y
e^{2y^2 } {\rm erfc}(\sqrt{2} y).
\end{equation}
Further, with $Z_{k,(N-k)/2}[u,u]$ denoting the canonical average of $\prod_{l=1}^k u(\lambda_l) \prod_{j=1}^{(N-k)/2}
u(x_j + i y_j)$ with respect to (\ref{3.1}), it has been shown in \cite{Si06} (taking $N$ even for
definiteness) that
\begin{eqnarray}\label{6.3}
&& \! \! Z_N[u,u]  :=\sum_{k=0}^N Z_{k,(N-k)/2}[u,u] 
\nonumber \\
& & \! \! = {2^{-N(N+1)/4} \over \prod_{l=1}^N \Gamma(l/2) }
{\rm Pf} [ \alpha_{j,k}(u) + \beta_{j,k}(u) ]_{j,k=1,\dots,N},
\end{eqnarray}
where, with $p_l(x)$ an arbitrary monic degree $l$ polynomial and $z:=x+iy$,
\begin{eqnarray}
&&\alpha_{j,k}(u)  =  \int_{-\infty}^\infty dx \, u(x)  \int_{-\infty}^\infty dy \, u(y) \nonumber \\
&& \qquad \times e^{-(x^2+y^2)/2}
p_{j-1}(x) p_{k-1}(y) {\rm sgn}(y-x) \label{6.4} 
\end{eqnarray}
\begin{eqnarray}
&&\beta_{j,k}(u)  =  
 2 i \int_{{\mathbb R}_+^2} dx dy \, u(z) e^{y^2 - x^2}
{\rm erfc} (\sqrt{2} y) \nonumber \\
&& \qquad \times  ( p_{j-1}(z) p_{k-1}(\bar{z}) - p_{k-1}(z) p_{j-1}(\bar{z})  ). \label{6.5}
\end{eqnarray}

It is the purpose of this Letter to report that all the results (\ref{6.1})-(\ref{6.3}) can be
generalized, thereby fully exhibiting the real Ginibre ensemble as exactly solvable.
For convenience it will be assumed throughout that $N$ is even. We first observe that with the
second argument $u$ on the LHS of (\ref{6.3}) replaced by an arbitrary
function $v=v(x,y)$ the equality remains valid with
$u$ in $\beta_{j,k}(u)$ replaced by $v$. With $p_{N,2n}$ denoting the probability that $2n$ out of the
$N$ eigenvalues are real, it then follows by choosing $v=1, u = \zeta$ that
\begin{eqnarray}\label{10}
&&\sum_{n=0}^{N/2} \zeta^n p_{N,2n} = {1 \over 2^{N(N+1)/4} \prod_{l=1}^N \Gamma(l/2) }
\nonumber \\
&& \qquad \times
{\rm Pf} [ \zeta \alpha_{j,k}(1) + \beta_{j,k}(1) ]_{j,k=1,\dots,N}
\end{eqnarray}
(cf.~Eq.~(11) of Ref.~\cite{KA05}).

As is well known in random matrix theory \cite{Me91,Fo02}, the correlations of a Pfaffian
generating functional (\ref{6.3}) are themselves Pfaffians. However in general this form 
involves the inverse of the matrix in (\ref{6.3}) with $u=1$. To make this explicit, one
seeks to choose the polynomials $\{p_l(x)\}$ to have the skew orthogonality property 
\begin{eqnarray}\label{pp1}
&&\! \!  \! \! \alpha_{2j,2k}(1) \! + \! \beta_{2j,2k}(1) \! = \! \alpha_{2j-1,2k-1}(1) \! + \! \beta_{2j-1,2k-1}(1) \!= \! 0, \nonumber \\
&& \! \!  \! \! 
\alpha_{2j-1,2k}(1) + \beta_{2j-1,2k}(1) = r_{j-1} \delta_{j,k}.
\end{eqnarray}
Our key result is that the very simple choice
\begin{eqnarray}\label{pp2}
&& p_{2j}(x) = x^{2j}, \quad p_{2j+1}(x) = x^{2j+1} - 2j x^{2j-1}, \nonumber \\
&& r_{j-1} = 2 \sqrt{2 \pi} \Gamma(2j-1)
\end{eqnarray}
validates (\ref{pp1}). With this established, and $q_{2j}(z) := - p_{2j+1}(z)$,
$q_{2j+1}(z) := p_{2j}(z)$, one finds for the correlations between complex eigenvalues
\begin{eqnarray}\label{8.a}&&
\! \! \rho_{(n)}^{\rm c}((x_1,y_1),\dots,(x_n,y_n)) = \prod_{j=1}^n \Big (2i e^{y_j^2 - x_j^2}
{\rm erfc}(\sqrt{2} y_j) \Big ) \nonumber \\ && \times
{\rm Pf} \left [ \begin{array}{cc} S^{\rm c}(\bar{z_j}, \bar{z}_k) &  S^{\rm c}(\bar{z_j}, {z}_k) \\
 S^{\rm c}({z_j}, \bar{z}_k) &  S^{\rm c}({z_j}, {z}_k) \end{array} \right ]_{j,k=1,\dots,n},
\end{eqnarray}
where $
S^{\rm c}(w,z) := \sum_{j=1}^N p_{j-1}(w) q_{j-1}(z) / r_{[(j-1)/2]} 
$ and $z_j := x_j + i y_j$. In the case $n=1$,  the Pfaffian equals $S^{\rm c}(\bar{z}_1,z_1)$ and
(\ref{6.2}) is reclaimed. In the case $n=2$ the Pfaffian equals
$
S^{\rm c}(\bar{z}_1,z_1) S^{\rm c}(\bar{z}_2,z_2) +
S^{\rm c}(\bar{z}_1,z_2) S^{\rm c}({z}_1,\bar{z}_2) -
S^{\rm c}(\bar{z}_1,\bar{z}_2) S^{\rm c}({z}_1,{z}_2).
$

Similarly, the correlations between real eigenvalues are computed as
\begin{eqnarray}\label{8.1}
&&\rho_{(n)}^{\rm r}(x_1,\dots,x_n) = \nonumber \\
&& \: \times 
{\rm Pf} \left [ \begin{array}{cc} {\rm sgn}(x_j-x_k) +  {I}^{\rm r}(x_j,x_k) & S^{\rm r}(x_j,x_k) \\
- S^{\rm r}(x_k,x_j) & D^{\rm r} (x_j,x_k)  \end{array} \right ]_{j,k=1,\dots,n}
\end{eqnarray}
with
$S^{\rm r}(x,y) = {1 \over 2} {\partial \over \partial y} I^{\rm r}(x,y)$, 
$D^{\rm r}(x,y) = {1 \over 2} {\partial \over \partial x} S^{\rm r}(x,y)$ and 
\begin{eqnarray}
 &&{I}^{\rm r}(x,y) =  \sqrt{2 \over \pi} e^{-x^2/2} \nonumber \\
 && \times 
\sum_{k=0}^{N/2-1} {x^{2k} \over (2k)!} \int_0^ye^{-u^2/2} u^{2k} \, du - (x \leftrightarrow y). \label{8.3}
\end{eqnarray}
In the case $n=1$ this gives $\rho_{(1)}(x) = S^{\rm r}(x,x) $ and (\ref{6.1}) is reclaimed. In the limit
$N \to \infty$ with $x,y$ fixed (\ref{8.3}) simplifies to
\begin{equation}\label{0.13a}
 {I}^{\rm r}(x,y) =  \sqrt{2 \over \pi} \int_0^{y-x} e^{-u^2/2} \, du
\end{equation}
implying the correlations decay at a Gaussian rate. Integrating (\ref{6.1}) over $x \in (-\infty,\infty)$ gives
the mean number $E_N$ of real eigenvalues, which is computed \cite{EKS94} to have the large $N$ asymptotic
form $\sqrt{2N/\pi} (1 - 3/8N - \dots)$. The variance $V_N$ of this same number is computed in terms of the
two-point correlation according to $V_N = \int_{-\infty}^\infty dx  \int_{-\infty}^\infty dy \,
\rho_{(2)}^{{\rm r} T}(x,y) + E_N$, $\rho_{(2)}^{{\rm r} T}(x,y) := \rho_{(2)}^{{\rm r} }(x,y)  -
\rho_{(1)}^{{\rm r} }(x) \rho_{(1)}^{{\rm r} }(y)$. 
We read off from (\ref{8.1}) that
\begin{eqnarray}
&& \rho_{(2)}^{{\rm r} T}(x_1,x_2) = - S^{\rm r}(x_1,x_2)  S^{\rm r}(x_2,x_1)  \nonumber \\
&& \qquad - \Big ({\rm sgn}(x_1 - x_2) + I^{\rm r}(x_1,x_2) \Big ) D^{\rm r}(x_1,x_2).
\end{eqnarray}
The quantity $\rho_{(2)}^{{\rm r} T}(x,y)/ \rho_{(1)}(x)$ is
integrable in $y$ showing that for large $N$, 
\begin{eqnarray}\label{0.13b}
&&V_N \sim E_N \Big (1 + \lim_{N \to \infty} (1/ \rho_{(1)}(0))
\int_{-\infty}^\infty \rho_{(2)}^{{\rm r} T}(0,y) \, dy \Big ) \nonumber \\
&& \quad = (2 - \sqrt{2}) E_N,
\end{eqnarray}
with the final equality
making use of (\ref{8.1})
and (\ref{0.13a}).

We draw attention to quantitatively similar results which hold for the zeros of the random
polynomial $p(z) = \sum_{j=0}^N \Big ( {N \atop j} \Big )^{1/2} \alpha_j z^j$, where
the $\alpha_j$ are i.i.d.~real Gaussian random variables. 
This has the interpretation in quantum mechanics as a random superposition of states
with spin $N/2$. Moreover, the function $p(e^{i \phi} \cot \theta/2)$ vanishes at the values
of $(\theta_j,\phi_j)$ on the sphere corresponding to the stereographic projection of the
zeros $z_j$ of $p(z)$, giving the Majorana parametrization \cite{Ha96}.
The analogy with the present problem is that
 the mean number of real
zeros is proportional to $\sqrt{N}$, as is the variance, and the correlations decay as Gaussians
\cite{BD97}. A distinction is the lack of a boundary for the eigenvalue support, which is
distributed as a Cauchy distribution. 

We remark too that although not reported on here,
the correlations between real and complex eigenvalues can be written as a Pfaffian.
Furthermore, we anticipate that the partially symmetric real Ginibre
ensemble, introduced in \cite{LS91}, will also yield to the present strategy. 

To leading order the support of the eigenvalue densities (\ref{6.1}), (\ref{6.2}) is the disk
$|z| = \sqrt{N}$, as is consistent with the formula (\ref{0.5})
for the boundary of the May stability criterion (here $c=1$ and $s=1$; 
however the variable $s$ can be reinstated by scaling $z \mapsto z/s$ throughout). Setting $x = \sqrt{N} + X$
and taking $N \to \infty$ in (\ref{6.1}) gives for the limiting edge profile
\begin{equation}\label{18}
\rho_{(1)}^{\rm r}(X) = {1 \over \sqrt{2 \pi} } \Big (
{1 \over 2} (1 - {\rm erf} \, \sqrt{2} X) + {e^{-X^2} \over 2 \sqrt{2} } (1 + {\rm erf} \, X) \Big ).
\end{equation}
For any fixed angle away from the real axis, as $N \to \infty$ the density of complex
eigenvalues near the boundary of support is radially symmetric, and the same as
that in the complex Ginibre ensemble. Writing the radius $r$ as $r = \sqrt{N} + R$,
for $N \to \infty$ this has the explicit form \cite{FH98,Ka05}
\begin{equation}
\rho_{(1)}^{\rm c}(R) = {1 \over \sqrt{2 \pi} } \Big ( 1 - {\rm erf} \, \sqrt{2} R \Big ),
\end{equation}
and is thus equal to twice the first term in (\ref{18}).

Suppose now that the variance of the Gaussian entries is reinstated as the variable $s^2$.
Let $R_r$ be the event there are no real eigenvalues, or
all real eigenvalues are less than $r$. By scaling of the
eigenvalues, 
Pr$(R_{s\sqrt{N}+sr})$ is independent of $s$, and for $N \to \infty$ it is an order 1 function
of $r$.  The latter
can be written as an infinite sum over the limiting $n$-point edge correlations, or
equivalently as a Fredholm determinant of the integral operator with kernel given by the edge
limit of the general entry in (\ref{8.1}). For $r$ large one has $\lim_{N \to \infty}{\rm Pr}(R_{s\sqrt{N}+sr}) 
\sim 1 - \int_r^\infty \rho_{(1)}^{\rm r}(X) \, dX$, showing that the corresponding PDF decays as
a Gaussian. For general $N$ a practical formula for computing this probability is in terms of the
generating functional (\ref{6.3}),
Pr$(R_{s\sqrt{N}+sr}) = Z_N[\chi_{\lambda \in (-\infty, s\sqrt{N} + sr)},1]$, where
$\chi_A=1$ if $A$ is true, $\chi_A=0$ otherwise, and
with the polynomials in (\ref{6.4}), (\ref{6.5}) chosen according to (\ref{pp2}). Numerical values of
Pr$(R_{s\sqrt{N}})$ so computed are tabulated in Table \ref{t1} for successive even values of $N$. The
quantity Pr$(\tilde{R}_{s \sqrt{N}}) := ({\rm Pr}(R_{s \sqrt{N}}) - p_{N,0})/(1 - p_{N,0})$, also
listed in Table \ref{t1}, gives the
probability that all real eigenvalues are less than
$s \sqrt{N}$, given that
there is at least one real eigenvalue. In the case $s^2 = 1/N$, this corresponds to the probability that all non-oscillatory solutions of
the linear system (\ref{0.2}) are stable, given that there is at least one non-oscillatory solution.

\begin{table}[th]
\begin{tabular}{|c||c|c|}\hline
$N$ & Pr$(R_{s\sqrt{N}})$ &  Pr$(\tilde{R}_{s\sqrt{N}})$ 
 \\[.1cm] \hline
$2$ & $0.81444$ & $0.737579$
\\[.1cm] \hline
$4$& $0.793864$ & $0.756706$ 
\\[.1cm] \hline
$6$ & $0.784485$ &  $0.762255$ 
   \\[.1cm]  \hline
$8$ & $0.778838$ &
$0.764193$  
\\[.1cm] \hline
$10$ & $0.774963$ & $0.76475$  \\[.1cm] \hline
$12$ & $0.772092$ & $0.76469$   \\[.1cm] \hline
$14$ & $0.769855$ & $0.76434$   \\[.1cm] \hline
$16$ & $0.768048$ & $0.76385$   \\[.1cm] \hline
\end{tabular}
\caption{\label{t1} Tabulation of two probabilities, specified in the text,
relating to the probability that all real eigenvalues of an $N \times N$
Gaussian real matrix, entries of mean zero, variance $s^2$, are less
that $s \sqrt{N}$. 
}
\end{table} 

With this study, building on the contributions of Lehmann and Sommers \cite{LS91}, Edelman \cite{Ed95},
Kanzieper and Akemann \cite{KA05} and Sinclair \cite{Si06}, the problem began by Ginibre over forty
years ago of calculating the statistical properties of the eigenvalues of non-Hermitian real Gaussian
matrices is solved. As a consequence the distribution of the largest real eigenvalue is presented in
a computable form. The largest real eigenvalue determines the stability of non-oscillatory solutions in
May's \cite{Ma72a} analysis of biological webs.

The work of PJF has been supported by the Australian Research Council.

\end{document}